\documentclass[11pt,superscriptaddress,aps,prd,preprint,nopacs,nofootinbib]{revtex4}
\usepackage[dvips]{graphicx}
\usepackage{amsfonts}
\usepackage{slashed}
\usepackage{color}

\usepackage{bm}
\usepackage{mathrsfs}
\usepackage{xcolor,color,graphicx,graphics}
\usepackage[all]{xy}
\usepackage{epsfig,subfigure}
\usepackage{latexsym,amssymb,amsmath,amsfonts} 
\usepackage[english]{babel} 
\usepackage[OT1]{fontenc}
\usepackage[latin1]{inputenc}
\usepackage{makeidx}
\usepackage{hyperref}
\usepackage{color,graphicx,graphics,wrapfig,epsf}

\newcommand{\be}{\begin{equation}}
\newcommand{\en}{\end{equation}}
\newcommand{\ba}{\begin{eqnarray}}
\newcommand{\ea}{\end{eqnarray}}
\newcommand{\bea}{\begin{eqnarray}}
\newcommand{\eea}{\end{eqnarray}}

\begin{document}

\title{Particle creation by wormholes: a $1+1$ model}

\author{Gonzalo Gurrea-Ysasi}

\address{Departamento de Construcciones Arquitect\'{o}nicas\\ Universidad Polit\'{e}cnica de Valencia, 46022 Valencia, Spain\\
gongurys@csa.upv.es}

\author{Gonzalo J. Olmo}

\address{Departament de F\'{i}sica Te\`{o}rica and IFIC, Centro Mixto Universitat de
Val\`{e}ncia - CSIC,\\
Universitat de Val\`{e}ncia, Burjassot-46100, Val\`{e}ncia, Spain\\ Departamento de F\'{\i}sica, Universidade Federal da 
Para\'{\i}ba,\\
 Caixa Postal 5008, 58051-970, Jo\~ao Pessoa, Para\'{\i}ba, Brazil\\
gonzalo.olmo@uv.es}

\begin{abstract}
The propagation of a free massless scalar field in a $1+1$ dimensional Minkowski space modeling a wormhole is considered. The wormhole model consists on two timelike trajectories, which represent the entrance and the exit of the wormhole, connected via some transfer function that specifies how incoming modes that reach the entrance are transferred to the exit. We find that particles and energy fluxes are generically produced except for transfer functions that represent global conformal transformations. We consider several examples involving exit trajectories which are asymptotically inertial, asymptotically null, and also involving a faster-than-light motion to illustrate the peculiarities of the emitted energy fluxes and quantum correlations. 
\end{abstract}


\maketitle

\tableofcontents

\section{Introduction}

The observation of gravitational waves opens a new window on the universe that will allow to put to a test predictions of General Relativity (GR) that have remained inaccessible for over a century \cite{TheLIGOScientific:2017first,TheLIGOScientific:2017qsa,Barack:2018yly} . Though the theory has been successful with all currently available astrophysical observations \cite{Will:2014kxa,Berti:2015itd,Akiyama:2019cqa}, there is still room for surprises. In particular, it is not yet possible to determine if the observed waves produced by the collision of compact objects really correspond to the classical black holes of GR or to some other exotic entities, such as boson stars \cite{Colpi:1986ye,Palenzuela:2017kcg,Cunha:2017wao}, gravastars \cite{Visser:2003ge,Pani:2010em,Chirenti:2016hzd}, or wormholes \cite{Cardoso:2016rao,Bambi:2013nla,Lobo:2007zb,Visser:1995cc,Morris:1988cz}, among others \cite{Ayon-Beato:2015eca}.
  
The case of wormholes is particularly attractive both from an objective theoretical perspective but also from the more subjective view  that corresponds to a non-scientific audience, as they feed our collective imagination very often led by science-fiction books and movies \cite{Sagan,Doraemon}. The theoretical and technological implications that the very existence of wormholes could have is enough to justify their theoretical scrutiny and a deeper analysis of their  physical implications. Aside from that, recent studies of extensions of Einstein's theory of general relativity (GR) indicate that the existence of wormholes might be rather natural. In fact, wormhole solutions generically arise in theories of the $f(R)$ type  \cite{Bambi:2015zch,Olmo:2015axa,Lobo:2009ip} and Born-Infeld type  \cite{Shaikh:2018yku,BeltranJimenez:2017doy,Olmo:2015dba,Shaikh:2015oha,Olmo:2013gqa,Harko:2013aya}, among others   \cite{Zubair:2016cde,Lobo:2014zla,Lobo:2014fma,Lobo:2013vga,Guendelman:2013sca,Lobo:2013prg,Harko:2013yb,Capozziello:2012hr,Dehghani:2009xu}, when coupled to standard matter sources such as electric fields  \cite{Afonso:2018mxn,Olmo:2013mla}, fluids \cite{Menchon:2017qed,Eiroa:2012nv}, and scalar fields of different types   \cite{Nascimento:2019qor,Afonso:2019fzv,Nascimento:2018sir,Afonso:2018bpv,Afonso:2017aci}. Some of those theories produce Schwarzschild-like solutions with an internal {\it black bounce} \cite{Carballo-Rubio:2019fnb,Simpson:2018tsi} which may turn into a traversable wormhole if the event horizon disappears by suitably tunning the parameters  \cite{Delhom:2019btt,Olmo:2015dba,Olmo:2013gqa}. In others, the Reissner-Nordstrom solution of GR gets modified by a wormhole of finite area that replaces the standard central point-like singularity \cite{Olmo:2015axa,Olmo:2012nx}. Rotating solutions with such internal wormhole structure are also possible \cite{OR_WiP}, and self-gravitating scalar field solutions yielding wormholes of astrophysical size have also been found \cite{Afonso:2019fzv,Afonso:2017aci}. 

A wormhole is typically viewed as an object of astrophysical nature that allows to connect two different universes or two distant regions within the same universe. Naturally, these distant regions could be separated in space, in time, or in both, which implies that wormholes could allow to implement the idea of time machine \cite{Visser:1995cc}. Given that quantum states are defined globally, the existence of wormholes must necessarily influence the properties of the vacuum state, which is the basic element on which particle states are built in a Fock quantization \cite{Birrell:1982ix}. This raises the question of the uniqueness and quantum stability of the vacuum state in geometries with wormholes. Therefore, in analogy with the case of black holes, one may wonder whether wormholes could lead to particle production. Will the vacuum state perceived by an observer on one side of the wormhole coincide with the vacuum state on the other side? If they do not coincide, what effects (number of particles, radiation fluxes, \ldots) will be felt by an observer crossing through the hole? These basic questions will be addressed in this work.  The close relation that one finds between wormholes and black holes also raises natural questions about the peculiarities of black hole evaporation in scenarios that involve wormholes \cite{Kuttner:2019tsm}. Rather than considering such an elaborate scenario, in this work we initiate a more modest programme in order to explore in a quantitative manner the effects of the  interaction of wormholes with quantum radiation fields, leaving aside for the moment any potential interactions between wormholes and event horizons. \\

A proper analysis of this problem requires a careful modelling of a wormhole, which can be done in different ways. Within GR, it is well known that one needs exotic energy sources, though this can be alleviated considering alternative theories of gravity. The classical estability of such solutions is also a problem of great relevance because it can cast doubts on the robustness and generality  of the conclusions derived from specific models. Thus, for the analysis of the quantum properties of wormholes it would be desirable to simplify  the classical geometric aspects to a minimum and retain only its key essential features. In this sense, we believe that the crucial defining characteristic of a wormhole is its ability to transport particles and energy from one space-time point (the entrance) to another (the exit), being all structural or engineering aspects of secondary relevance. For this reason, we will consider point-like structureless wormholes in 1+1 dimensions\footnote{Similar simplifications were considered in Chapter 18 of \cite{Visser:1995cc}.}. This drastic simplification will allow us to focus on the quantum aspects of the model in a scenario where the technical aspects of the quantization are also easier to handle, which will facilitate the quantitative analysis of the properties of the radiation fields. We will thus model a wormhole as a device consisting of two points  separated  in space-time, one defining the entrance and the other the exit, connected via some transfer function. The transfer function is necessary in order to specify how the modes of the quantum field that reach the entrance, which follows a certain trajectory in space-time, are transferred to a specific point on the space-time trajectory of the exit. As we will see, this transfer function will play a central role in the determination of the radiation properties of the quantum field. \\

Our 1+1 wormhole model is conceptually closely related to the moving mirror models studied by Moore \cite{Moore1970}, DeWitt \cite{DeWitt1975}, Fulling and Davies \cite{Fulling_Davies1976} back in the seventies of the past century (see also Refs.~\cite{Birrell:1982ix,Crispino:2007eb}).  The moving mirror example is very helpful to understand quantum radiation problems and mimick the properties of black hole evaporation in a Minkowskian scenario. In this setup,  a field (typically a scalar) is forced to vanish at the surface of the mirror, being this surface a moving boundary in 1+1 Minkowski space. Depending on the mirror trajectory, quanta can be created, giving rise to a process of particle creation analogous to that occurring in  black hole formation-evaporation scenarios. In our wormhole case, the vanishing boundary condition at the mirror location (total reflection) is replaced by the identification of the field mode at specific points of the trajectories of the wormhole entrance and exit (total transmission). Since the entrance and exit are represented by curves in space-time (see Fig.\ref{fig:1}), a transfer function must be specified to determine how the identification of points on $\gamma_1$ (entrance) and $\gamma_2$ (exit) proceeds. We will see that, in general, the existence of a wormhole implies the lack of a unique notion of vacuum state, which leads to the phenomenon of particle creation. The amount of created particles and the intensity of its associated energy fluxes will be analyzed in detail, paying special attention to those cases in which an unbounded emission can occur. \\

The paper is organized as follows. In Sec.\ref{sec:II} we briefly review the quantization of a massless scalar field in $1+1$ dimensions, where we introduce the basic computational tools to be used in the rest of the paper. In Sec.\ref{sec:III} we describe our wormhole setup and how to obtain the field modes out of this model. Sec. \ref{sec:IV} presents several examples of wormholes and transfer functions and their physical implications. We conclude the paper with a summary and discussion.

\section{Basics of quantum fields in 1+1 Minkowski space} \label{sec:II}

In order to introduce the mathematical tools we will use for the analysis of quantum radiation problems in 1+1 dimensions, we will  consider a massless scalar field as an illustration. The scalar satisfies the equation
\begin{equation}
\Box\phi(t,z)=0 
\end{equation}
which in double null coordinates $x^\pm=t \pm z$ becomes
\begin{equation}\label{eq:fieldeq}
\partial_-\partial_+\phi(x^-,x^+)=0  \ .
\end{equation}
The general solution to this equation can be written as 
\begin{equation}
u(x^+,x^-)=u_{L,i}(x^+)+u_{R,i}(x^-)=u_{L,i}(t+z)+u_{R,i}(t-z) \ ,
\end{equation}
which shows that generic modes are a superposition of right-moving and left-moving waves, being the right-moving and left-moving sectors independent if no boundaries are imposed. As is evident, the subindices $R$ and $L$ denote left-moving and right-moving quantities, respectively.  The modes are normalized according to the scalar product
\begin{equation}\label{eq:sprod}
(u_1, u_2)=-i\int_\Sigma d\Sigma^\mu (u_1 \partial_{\mu}u_2^*-u_2^*\partial_{\mu} u_1),
\end{equation}
where $\Sigma$ is an arbitrary Cauchy hypersurface.  In two dimensions we can take advantage of the null coordinates to write this scalar product in simplified form as
\begin{equation}\label{eq:sprodsimp}
 (u_1, u_2)=-2i\int^{+\infty}_{-\infty} dx^\pm u_1 \partial_{x^\pm}u_2^*
\end{equation}
where an integration by parts has been performed assuming that the modes decay sufficiently fast at infinity, which is well-justified for localized wave-packets. The quantum field can thus be written as a linear superposition of the form
\begin{equation}
\phi(x^+,x^-)=\sum_i \left(a_{L,i} u_{L,i}(x^+)+a_{L,i}^\dagger u_{L,i}^*(x^+)+a_{R,i} u_{R,i}(x^-)+a_{R,i}^\dagger u_{R,i}^*(x^-) \right) \ ,
\end{equation}
where the creation and anihilation operators satisfy the usual commutation relations
\begin{equation}
[a_{L,i},a_{L,j}^\dagger]=\delta_{ij}\hbar \ , \ [a_{R,i},a_{R,j}^\dagger]=\delta_{ij}\hbar
\end{equation}
with all other commutators vanishing. 

The vacuum state $\left|0\right \rangle_x$ is thus defined as the state anihilated by the $a_{L,i}$ and $a_{R,i}$ operators 
\begin{equation}
a_{L,i}\left|0\right \rangle_x=0 \ , \ a_{R,i}\left|0\right \rangle_x=0
\end{equation}
Given that (\ref{eq:fieldeq}) is invariant under conformal transformations $y_\pm=y_\pm(x_\pm)$, the field could also be decomposed in a different set of modes of the form 
\begin{equation}
\phi(y^+,y^-)=\sum_i \left(b_{L,i} v_{L,i}(y^+)+b_{L,i}^\dagger v_{L,i}^*(y^+)+b_{R,i} v_{R,i}(y^-)+b_{R,i}^\dagger v_{R,i}^*(y^-) \right) \ , 
\end{equation}
with a new set of creation and anihilation operators 
\begin{equation}
[b_{L,i},b_{L,j}^\dagger]=\delta_{ij}\hbar \ , \ [b_{R,i},b_{R,j}^\dagger]=\delta_{ij}\hbar
\end{equation}
with all other commutators vanishing. The corresponding vacuum state $\left|0\right \rangle_y$ is determined by 
\begin{equation}
b_{L,i}\left|0\right \rangle_y=0 \ , \ b_{R,i}\left|0\right \rangle_y=0
\end{equation}
As both sets of modes are complete, the two bases can be related via the so-called Bogolubov transformations
\begin{eqnarray}
v_{L,j}(y^+)&=&\sum_i \left(\alpha_{ji}v_{L,i}(x^+)+\beta_{ji}u_{L,i}^*(x^+)\right) \\
v_{R,j}(y^-)&=&\sum_i \left(\gamma_{ji}u_{R,i}(x^-)+\eta_{ji}u_{R,i}^*(x^-)\right) 
\end{eqnarray}
where the coefficients are determined by the scalar products
\begin{eqnarray}
\alpha_{ji}=(v_{L,j},u_{L,i}) \ & , & \ \beta_{ji}=-(v_{L,j},u_{L,i}^*) \\
\gamma_{ji}= (v_{R,j},u_{R,i}) \ & , & \ \eta_{ji}= -(v_{R,j},u_{R,i}^*)\\
\end{eqnarray}
and lead to a linear relation between creation and anihilation operators of the form
\begin{equation}
b_{L,j}=\sum_i\left(\alpha^*_{ji}a_{L,i} -\beta_{ji}a_{L,i}^\dagger\right) \ , \ b_{R,j}=\sum_i\left(\gamma^*_{ji} a_{R,i}-\eta_{ji}a_{R,i}^\dagger\right) \ ,
\end{equation}
from which expressions for $b_{R,j}^\dagger$ and $b_{L,j}^\dagger$ can be derived. From this relations it is immediate to see that the expectation value of the number operator $N^y_{L,j}=b^\dagger_{L,j}b_{L,j}$ corresponding to the expansion in  coordinates  $y_\pm$ evaluated on the vacuum of the $x_\pm$ observer is given by 
\begin{equation}
_{x}\langle 0|N^y_{L,i}|0\rangle_{x}=\sum_k\left |\beta_{ik}\right |^2   
\end{equation}
The definitions and results derived so far are all standard textbook material (see, for instance, Refs.~\cite{Birrell:1982ix,Parker:2009uva}). Next we will introduce some less known material first presented in Ref.~\cite{Fabbri:2004yy}, further developed in Ref.~\cite{OlmoThesis}, and applied in  Refs.~\cite{Agullo:2010hi,Agullo:2009vq,Agullo:2008qb,Agullo:2006um}. 
From the definition of $\beta_{ik}$ in terms of the scalar product (\ref{eq:sprod}) we can take advantage of the simplicity of the two-dimensional model to obtain an expression for the number of particles in terms of the two-point correlation functions of the field. To see this, note that using (\ref{eq:sprodsimp}) we can write the expectation values $_{x}\langle 0|b^\dagger_{R,i}b_{R,j}|0\rangle_{x}$ and $_{x}\langle 0|b^\dagger_{L,i}b_{L,j}|0\rangle_{x}$ as
\begin{equation}
_{x}\langle 0|b_{i}^\dagger b_{j}|0\rangle_{x} =-4 \sum_k \int_{-\infty}^{+\infty}  \int_{-\infty}^{+\infty} dy_1 dy_2 v_{i}(y_1)v_{j}^*(y_2) \partial_{y_1}u_{k}^*(x_1)  \partial_{y_2}u_{k}(x_2) \ ,
\end{equation}
where for notational simplicity we have omitted the index $R/L$ when the integration is performed over the variable $y^-/y^+$. Simple manipulations bring the above equation into 
  \begin{equation}
_{x}\langle 0|b_{i}^\dagger b_{j}|0\rangle_{x} =-4 \int_{-\infty}^{+\infty}  \int_{-\infty}^{+\infty} dy_1 dy_2  v_{i}(y_1)v_{j}^*(y_2) \frac{dx_1}{dy_1}\frac{dx_2}{dy_2}\sum_k \partial_{x_1}u_{k}^*(x_1)  \partial_{x_2}u_{k}(x_2) \ , 
\end{equation}
which can also be written as 
  \begin{equation}\label{eq:bbx0}
_{x}\langle 0|b_{i}^\dagger b_{j}|0\rangle_{x} =-4\int_{-\infty}^{+\infty}  \int_{-\infty}^{+\infty} dy_1 dy_2  v_{i}(y_1)v_{j}^*(y_2) \frac{dx_1}{dy_1}\frac{dx_2}{dy_2} \  _{x}\langle 0|\partial_{x_1}\phi(x_1)  \partial_{x_2}\phi(x_2)|0\rangle_{x} \ 
\end{equation}
because 
  \begin{equation}
_{x}\langle 0|\partial_{x_1}\phi(x_1)  \partial_{x_2}\phi(x_2)|0\rangle_{x} =\sum_k \partial_{x_1}u_{k}^*(x_1)  \partial_{x_2}u_{k}(x_2) \ .
\end{equation}
The explicit form of this two-point function can be obtained by direct computation or via symmetry arguments \cite{DiFrancesco:1997nk,Ginsparg:1988ui}. By direct computation, it is convenient to consider plane wave modes of the form $u_\omega(x)=\frac{e^{-i\omega x}}{\sqrt{4\pi\omega}}$ such that 
  \begin{equation}
_{x}\langle 0|\partial_{x_1}\phi(x_1)  \partial_{x_2}\phi(x_2)|0\rangle_{x} =\int_0^\infty d\omega \frac{\omega}{4\pi} e^{i\omega (x_1-x_2)}=-\frac{i}{4\pi}\frac{\partial}{\partial \Delta}  \int_0^\infty d\omega e^{i\omega \Delta } \ .
\end{equation}
Though this integral diverges in the ultraviolet, one can regularize it by replacing the term $\Delta\equiv x_1-x_2$ by $\Delta+i\epsilon$ in the exponential and then taking the limit $\epsilon\to 0$ at the end. One then finds that  
\begin{equation}
_{x}\langle 0|\partial_{x_1}\phi(x_1)  \partial_{x_2}\phi(x_2)|0\rangle_{x}=-\frac{1}{4\pi}\frac{1}{(x_1-x_2)^2} \ .
\end{equation}
A more convenient expression for Eq. (\ref{eq:bbx0}) can be obtained by normal ordering the two point function via the subtraction of  $_{y}\langle 0|b_{i}^\dagger b_{j}|0\rangle_{y}$, which should vanish by construction. One thus finds that  the above expectation value can be written as
  \begin{equation}\label{eq:bbx}
_{x}\langle 0|b_{i}^\dagger b_{j}|0\rangle_{x} =\frac{1}{\pi} \int_{-\infty}^{+\infty}  \int_{-\infty}^{+\infty} dy_1 dy_2  v_{i}(y_1)v_{j}^*(y_2)\left[ \frac{dx_1}{dy_1}\frac{dx_2}{dy_2}\frac{1}{(x_1(y_1)-x_2(y_2))^2}-\frac{1}{(y_1-y_2)^2} \right]\ .
\end{equation}
This formula will play an important role in this paper. It puts forward that the expectation value of the number operator is given by  the diagonal elements of $_{x}\langle 0|b_{i}^\dagger b_{j}|0\rangle_{x}$, which can be computed by projecting with the modes $v_{i}(y)$ and $v_{j}^*(y)$ of the $y^\pm$ basis on a (geometric) kernel that contains all the information of the quantum field. It is important to note that this kernel is invariant under Poincar\'{e} transformations (space-time shifts, $x\to y+y_0$, plus Lorentz boosts, $x\to \gamma y$, with $y_0$ and $\gamma$ some constants) and also special conformal transformations (trajectories with constant proper acceleration), which are contained in the so-called Mobius transformations $x(y)=\frac{ay+b}{cy+d}$, with $ad-bc\neq 0$. As a result, observers related by such coordinate transformations share the same vacuum and have a vanishing $_{x}\langle 0|b_{i}^\dagger b_{j}|0\rangle_{x}$. For any other type of conformal relations, the kernel will not vanish and particles could, in principle, be observed. The observatility of such particles depends on the specific choice of modes, as the outcome of the integral depends crucially on them. In this sense, if one considers localized wave-packets, the integral will only yield a non-vanishing result whenever the difference of correlations between $_{x}\langle 0|\partial_{x_1}\phi(x_1)  \partial_{x_2}\phi(x_2)|0\rangle_{x}$ and $_{y}\langle 0|\partial_{y_1}\phi(y_1)  \partial_{y_2}\phi(y_2)|0\rangle_{y}$ is not zero on the region that supports the wave-packet. Note that for such localized wave-packets, the main contribution will typically come from nearby points $y_1\to y_2$. An expansion of the two-point correlators around coincident points leads to 
 \begin{equation}
-\frac{1}{4\pi}\lim_{y_2\to y_1}\left[ \frac{dx_1}{dy_1}\frac{dx_2}{dy_2}\frac{1}{(x_1(y_1)-x_2(y_2))^2}-\frac{1}{(y_1-y_2)^2} \right]=-\frac{1}{24\pi}\left\{\frac{x^{'''}(y_1)}{x^{'}(y_1)}-\frac{3}{2}\frac{(x^{''}(y_1))^2}{(x^{'}(y_1))^2} \right\} +O(y_1-y_2)\ ,
\end{equation}
where the right-hand side represents the so-called Schwarzian derivative, typically denoted as $-\frac{1}{24\pi}\{x,y\}$, and represents the anomalous transformation under conformal changes of coordinates of the normal-ordered stress energy tensor (normal-ordering breaks diffeomorphism invariance) 
\begin{equation}
\langle \Psi|:T_{\pm\pm}(y^\pm):|\Psi\rangle = \left(\frac{dx^\pm}{dy^\pm}\right)^2\ \langle \Psi|:T_{\pm\pm}(x^\pm):|\Psi\rangle-\frac{1}{24\pi}\{x^\pm,y^\pm\} \ .
\end{equation}
Since (\ref{eq:bbx}) is evaluated in the vacuum state of the observer $x$, $|\Psi\rangle=|0\rangle_{x}$,  it follows that $_{x}\langle 0|:T_{\pm\pm}(y^\pm):|0\rangle_{x}=-\frac{1}{24\pi}\{x^\pm,y^\pm\} $ represents the energy flux observed by the observer $y$ on the vacuum state of $x$. Given that $\{x,y\}$ only vanishes for Mobius transformations, any transformation that breaks global conformal invariance (the symmetry group of the vacuum state) will lead to the emission of energy fluxes and the creation of particles \cite{Fabbri:2004yy}. The main virtue of Eq.(\ref{eq:bbx}) is that it will allow us to visualize in a straightforward manner the production of particles and the emission of energy fluxes by just considering the use of localized wavepackets. In this sense, it represents a clear advantage over the traditional {\it black box} approach in terms of Bogolubov coefficients.

\section{Modeling wormholes in 2D Minkowski space-time} \label{sec:III}

\begin{figure}[h]
\centering
\includegraphics[width=0.60\textwidth]{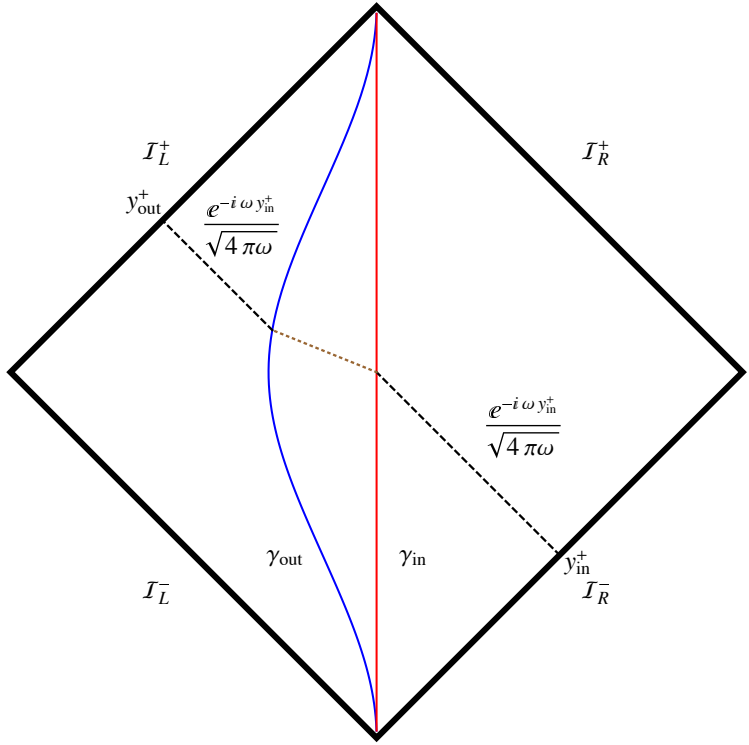}
\caption{Effect of a wormhole on a plane wave in 1+1 Minkowski space. The curve $\gamma_{in}$ represents the entrance, which is located at $z=0$, while the curve $\gamma_{out}$ represents the exit, located at $z=-1/2$.   \label{fig:1}}
\end{figure}

As mentioned in the introduction, in our view the key defining characteristic of a wormhole is its ability to transport particles and information from one point to another without affecting their properties. In this respect, the illustration in Fig.\ref{fig:1} captures this idea by representing a plane wave that departs from $\mathcal{I}^-_R$ moving along a given $y^+_{in}=$constant ray and upon reaching the curve $\gamma_1$ it is transfered without alteration to a point on the curve $\gamma_2$ at the location $y^+_{out}$, from which it continues propagating following a constant $y^+_{out}$ trajectory up to $\mathcal{I}^+_L$. If we represent the field mode as a plane wave that departs from $y^+_{in}$ at $\mathcal{I}^-_R$ with phase $\omega y^+_{in}$,  it will reach $\mathcal{I}^+_L$ with the same phase but following a new null geodesic characterized by $y^+_{out}=\tau(y^+_{in})$. We will refer to the function $\tau$ as the transfer function. Essentially, this function shifts the incoming plane wave along the $y^+$ axis from its original $y^+_{in}$ geodesic to its final $y^+_{out}$ geodesic.

\begin{figure}[h]
\centering
\includegraphics[width=0.60\textwidth]{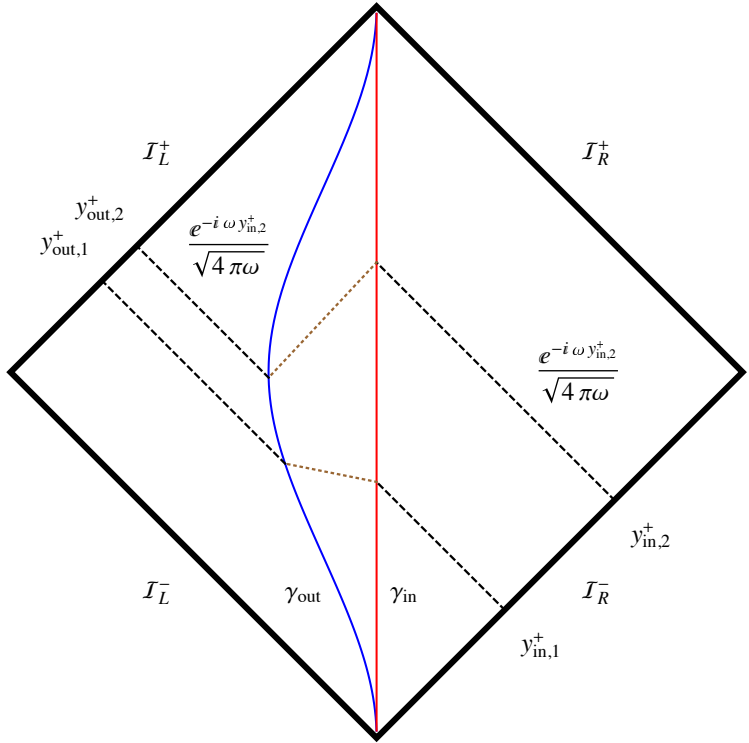}
\caption{Propagation of two different modes to illustrate that the $y^+$ dependence of a  mode along a given $y^+$ geodesic is, in fact, made out of two different pieces. Note that the numerical value of $y^+_{in,1}$ has been chosen to coincide with $y^+_{out,2}$ so as to align those two rays along the same $y^+=$constant geodesic. \label{fig:2}}
\end{figure}
\begin{figure}[h]
\centering
\includegraphics[width=0.60\textwidth]{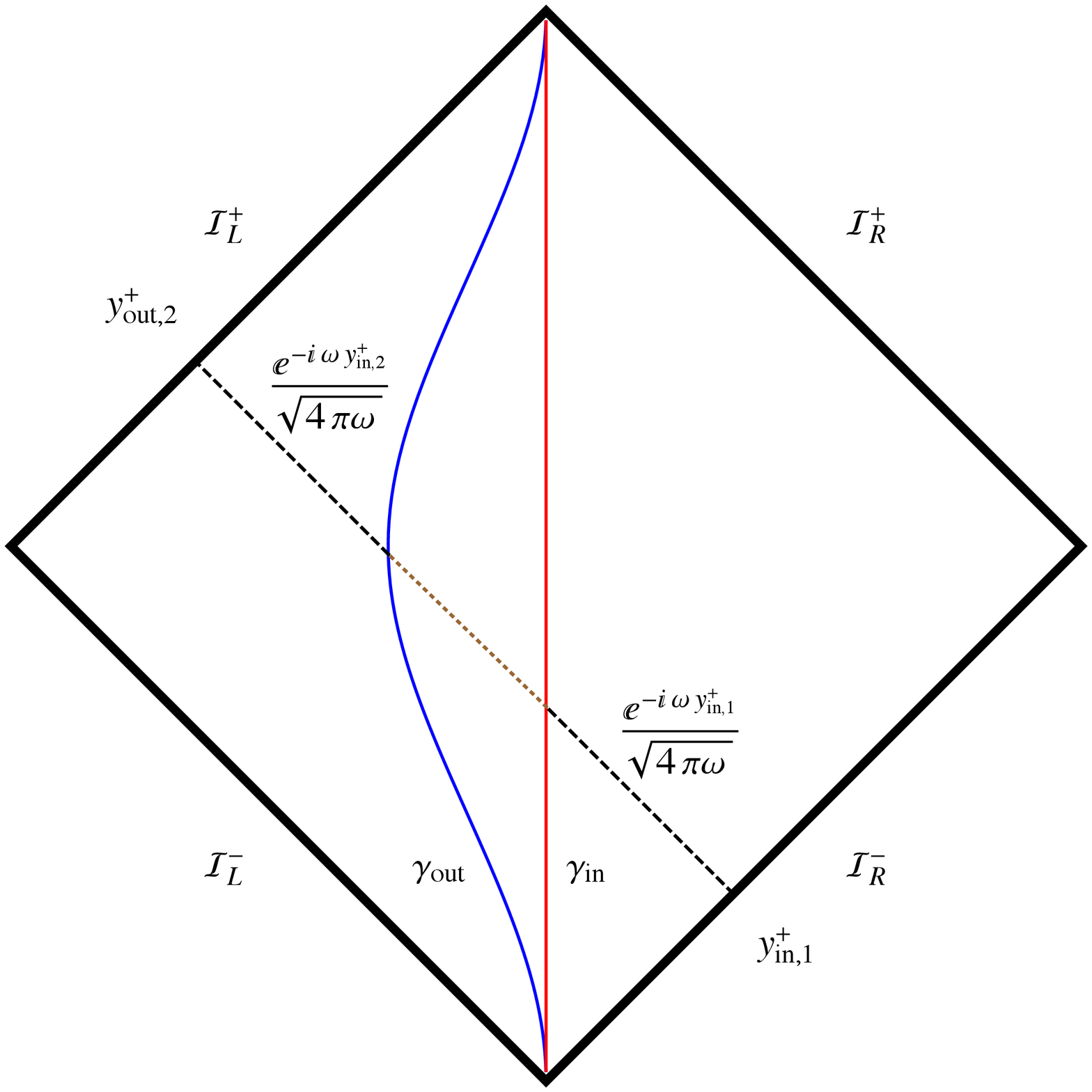}
\caption{The plane wave that reaches $\mathcal{I}^+_L$ along $y^+_{in,1}$ originated at  $\mathcal{I}^-_R$ at $y^+_{in,2}$, as shown in Fig.\ref{fig:2}. The transfer function then shifted the incoming geodesic initially propagating along $y^+_{in,2}$ to the final trajectory $y^+_{out,2}=y^+_{in,1}$. \label{fig:3}}
\end{figure}
Once the transfer function has been defined, it is possible to determine the explicit dependence of the left-moving field modes on $y^+$ and $y^-$ with the help of Figs. \ref{fig:2} and \ref{fig:3}. Similar considerations apply to the right-moving modes though, in principle, a different transfer function could apply, depending on the wormhole designer's choice. 
The key point is to determine the form of a mode along a geodesic with $y^+=$constant that extends from $\mathcal{I}^-_R$ to $\mathcal{I}^+_L$. As illustrated with the rays traced on those Penrose diagrams, below the curve $\gamma_{in}$ the mode along a given $y^+_{in,1}$ is determined by a wave with phase $\omega y^+_{in,1}$. However, the propagation of the null geodesic beyond $\gamma_{out}$ is, in general, followed by a modification of the wave phase, which is now coming from a different value of $y^+_{in}$, such that the new phase is $\omega y^+_{in,2}$. In the particular case in which $y^+_{out,2}=\tau(y^+_{in,2})=y^+_{in,1}$, we have that $y^+_{in,2}=\tau^{-1}(y^+_{in,1})$, which is the situation depicted in Fig. \ref{fig:3}. This allows us to write the modes that propagate from $\mathcal{I}^-_R$ to $\mathcal{I}^+_L$ as
\begin{equation}
\phi^{in}_\omega(y^+,y^-)=\frac{e^{-i\omega X^+(y^+)}}{\sqrt{4\pi\omega}} \ \text{ where } X^+(y^+)= \left\{\begin{array}{lr}
y^+ & \text{ if } y^-<\gamma_{in}(y^+) \\
\ \\
 \tau^{-1}(y^+) & \text{ if } y^->\gamma_{out}(y^+)
\end{array}\right. \ ,
\end{equation}
  where there is no need to specify the subindex $in$ or $out$ because both coordinates range from $-\infty$ to $+\infty$. The creation and anihilation operators associated to the modes $\phi^{in}_\omega(y^+,y^-)$ define the natural vacuum on $\mathcal{I}^-_R$ and will be denoted as $|0\rangle_X$. A similar decomposition can be performed by considering the backwards propagation of modes at $\mathcal{I}^+_L$. In this case, along a null geodesic with $y^+_{out,2}=$constant we have a mode with phase $\omega y^+_{out,2}$ at $\mathcal{I}^+_L$ which propagated backwards reaches $\gamma_{out}$ and comes out on the other side of $\gamma_{in}$ with a phase $\omega y^+_{out,2}=\omega \tau(y^+_{in,2})$. However, the mode that propagates along $y^+_{out,2}$ below  $\gamma_{in}$ originated at a $y^+_{out,1}$ with a phase $\omega y^+_{out,1}=\omega \tau(y^+_{in,1})$, and given that the numerical value of this $y^+_{in,1}$ coincides with $y^+_{out,2}$, we can conclude that in the out region the field admits a mode decomposition of the form 
 \begin{equation}
\phi^{out}_\omega(y^+,y^-)=\frac{e^{-i\omega Y^+(y^+)}}{\sqrt{4\pi\omega}} \ \text{ where } Y^+(y^+)= \left\{\begin{array}{lr}
y^+ & \text{ if } y^->\gamma_{out}(y^+) \\
\ \\
 \tau(y^+) & \text{ if } y^-<\gamma_{in}(y^+)
\end{array}\right. \ .
\end{equation}
 The creation and anihilation operators associated to the modes $\phi^{out}_\omega(y^+,y^-)$ define the natural vacuum on $\mathcal{I}^+_L$ and will be denoted as $|0\rangle_Y$. 
 \begin{figure}[h]
\centering
\includegraphics[width=0.60\textwidth]{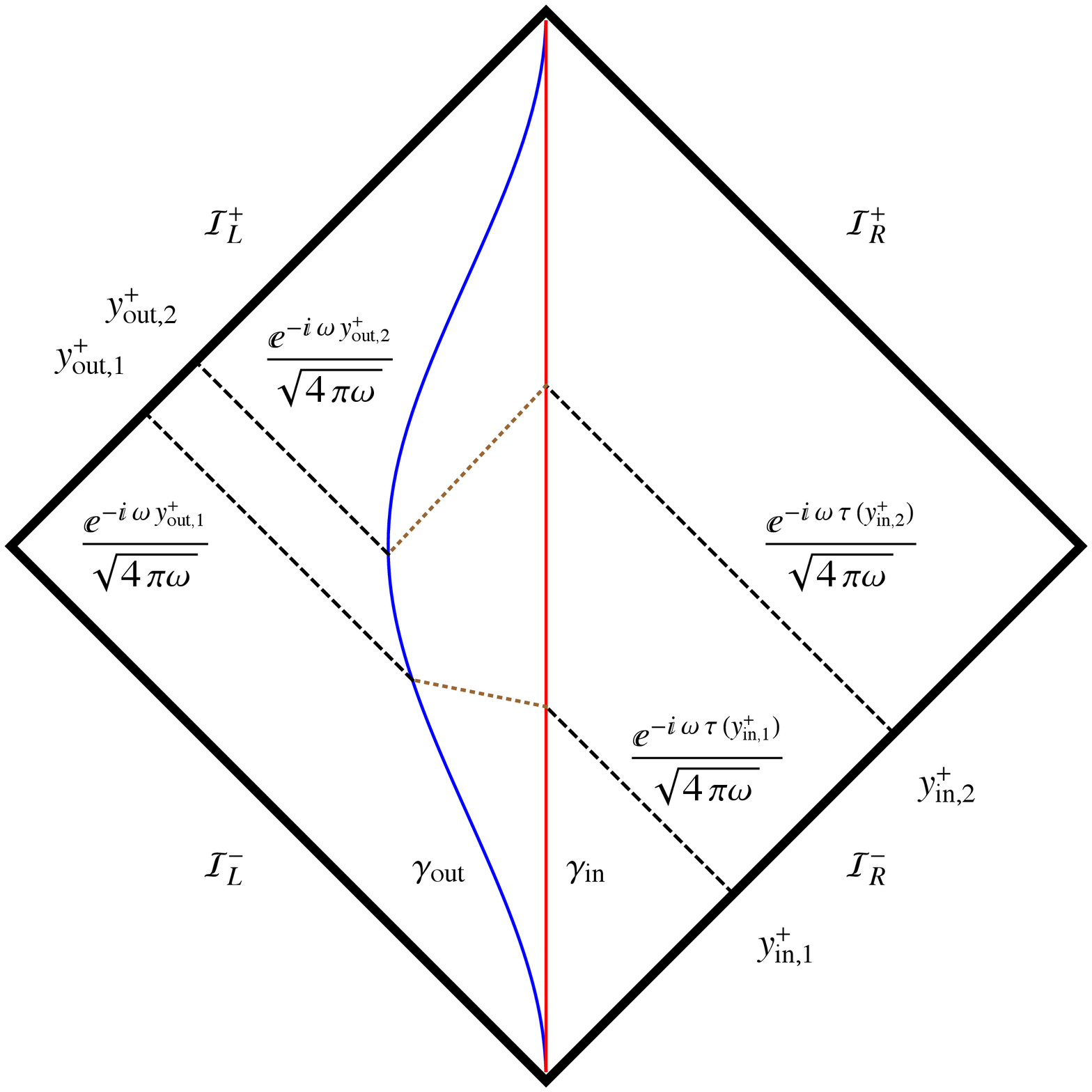}
\caption{Backwards propagation of the {\it out} modes. Note that the mode along $y^+_{in,1}=y^+_{out,2}$ is made out of two pieces, like in Figs.\ref{fig:2} and \ref{fig:3}. \label{fig:4}}
\end{figure}

 \section{Particles and energy fluxes}\label{sec:IV}

We are now ready to analyze the properties of the number operator applying the general methodology of Section \ref{sec:II} to the wormhole scenario presented above. For this purpose, we consider the expansion of the field in the {\it in} and {\it out} bases in such a way that the expectation value $_{X}\langle 0|b_{L,\omega}^\dagger b_{L,\omega'}|0\rangle_{X} $ now takes the form 
 \begin{equation}\label{eq:bbxWH}
_{X}\langle 0|b_i^\dagger b_j|0\rangle_{X} =\frac{1}{\pi}\int  \int_{\mathcal{I}^+_L} dy_1 dy_2  v_i(y_1)v_j^*(y_2)\left[ \frac{dX^+_1}{dy_1}\frac{dX^+_2}{dy_2}\frac{1}{(X^+_1(y_1)-X^+_2(y_2))^2}-\frac{1}{(y_1-y_2)^2} \right] \ , 
\end{equation}
where in the case of plane waves the subindices $i$ and $j$ stand for the pair $\{L,\omega\}$, such that $v_i(y)\to v_{L,\omega}(y)=\frac{e^{-i\omega y}}{\sqrt{4\pi \omega}}$.  Given that we are performing the integral over the {\it out} coordinate $y^+$ on $\mathcal{I}^+_L$, which corresponds to $y^-\to +\infty$, it follows that $X^+(y^+)=\tau^{-1}(y^+)$ is independent of the particular boundaries $\gamma_{in}$ and $\gamma_{out}$ that define the trajectories of the wormhole entrance and exit. The only relevant information about the wormhole is provided by the inverse of the transfer function $\tau(y^+)$.  Obviously this is an idealized situation in which the incoming wave is perfectly transmitted through the hole. More {\it realistic} models could take into account the effect of partial transmission as well as backscattering, which is likely to introduce some dependence on the $\gamma_{in}$ and $\gamma_{out}$ trajectories. Here we will stick to the simplest scenario in which such refinements are neglected.  \\
It is worth noting that the integral (\ref{eq:bbxWH}) can also be performed over the variable $X^+$. Given that $y=\tau(X^+)$, $X^+$ can be seen as the {\it in } coordinate. After elementary manipulations, the expectation value (\ref{eq:bbxWH}) can be written as 
\begin{equation}\label{eq:bbxWH2}
_{X}\langle 0|b_i^\dagger b_j|0\rangle_{X} =\frac{1}{\pi}\int  \int_{-\infty}^{\infty} dX_1 dX_2  v_i(\tau(X_1))v_j^*(\tau(X_2))\left[ \frac{1}{(X_1-X_2)^2}-\frac{\tau_{X_1}\tau_{X_2}}{(\tau(X_1)-\tau(X_2))^2} \right] \ , 
\end{equation}
where the superindex in $X$ has been omitted for simplicity. In (\ref{eq:bbxWH}), the integrand is directly related to $_{X}\langle 0|:\partial_{y_1}\phi  \partial_{y_2}\phi:|0\rangle_{X}$ projected with the natural {\it out} modes on $\mathcal{I}^+_L$, while in (\ref{eq:bbxWH2}) it depends on the {\it out} modes propagated backwards to $\mathcal{I}^-_R$ and multiplied by $-_{y}\langle 0|:\partial_{X_1}\phi  \partial_{X_2}\phi:|0\rangle_{y}$. Another useful alternative, is to keep the integration over the {\it out} $y$ coordinate and use that $y=\tau(X)$ and $X_y=1/\tau_X$ to rewrite the normal ordered correlator in  (\ref{eq:bbxWH}) as
\begin{equation}\label{eq:bbxWH3}
_{X}\langle 0|:\partial_{y_1}\phi  \partial_{y_2}\phi:|0\rangle_{X} =-\frac{1}{4\pi}\left[ \frac{1}{\tau_{X_1}\tau_{X_2}}\frac{1}{(X_1-X_2)^2}-\frac{1}{(\tau(X_1)-\tau(X_2))^2} \right] \ .
\end{equation}
This form allows for a straightforward evaluation and graphic representation of this quantity in parametric form, and will be used later on in our discussions.

In the following we will consider different transfer functions and will discuss their implications for particle production and energy fluxes. 

\subsection{Example I: identity, boosts, and Mobius transfer functions.}

Let us begin by considering any two trajectories $\gamma_{in}(y^+)$ and $\gamma_{out}(y^+)$ together with some transfer function $\tau(y^+)$ for the left-moving modes. For the particular case in which $\tau(y^+)=y^+$, the wormhole simply transfers any incoming wave from $\gamma_{in}(y^+)$ to $\gamma_{out}(y^+)$ without altering the value of $y^+$. The wormhole is certainly operating, transferring incoming waves on the entrance $\gamma_{in}(y^+)$ to the exit $\gamma_{out}(y^+)$, thus making the wave to go from some $y^-_{in}=\gamma_{in}(y^+)$ to $y^-_{out}=\gamma_{out}(y^+)$ instantaneously, which could be seen as a jump into the future\footnote{Actually, since this transfer occurs along $y^+=$constant, the jump in time is $\Delta t=\Delta y^-/2$ and is accompanied by a spatial jump  $\Delta z=-\Delta y^-/2$}. Evaluation of $_{X}\langle 0|b_i^\dagger b_j|0\rangle_{X}$ for this transfer function yields trivially a vanishing result because the difference between correlators vanishes everywhere. And the same happens for any $\tau(y^+)$ of the Mobius type, which include boosts, constant shifts, and constant proper acceleration trajectories. For such scenarios no particles or energy fluxes are detected at $\mathcal{I}^+_L$. 

Note that this is so because the left-moving modes do not depend on $y^-$  and, for this reason, they are insensitive\footnote{ In a non-perfectly transmitting wormhole, one expects a dependence of the modes on $y^-$ due to the reflected and backscattered part of the modes, which would make the local correlations before and after crossing the hole different, potentially causing some particle creation.} to the shifts in the $y^-$ coordinate induced by the wormhole when trasferring the modes from  $\gamma_{in}(y^+)$ to $\gamma_{out}(y^+)$. The lack of particle production for the identity, boosts ($y \to a y$), and Poincar\'e transformations is easy to understand because the change in the phase of the field modes is trivial in the first case, can be reabsorbed into a constant redefinition of the frequency in the second (Doppler shift), or simply amounts to a redefinition of the origin of coordinates in the third. The case of full Mobius transformations is not so easy to visualize and has led to some confusion in the literature in the past \cite{Fabbri:2004yy,Birrell:1982ix}. 

If the transfer function for the right-moving modes had the same property, namely, that $\tau_R(y^-)=y^-$ (or any other Mobius form), then no particles would be detected at $\mathcal{I}^+_R$ either. In practical terms, there would be no difference with respect to a standard Minkowski space despite having a wormhole. In other words, the vacuum state would be the same as in Minkowski space. \\ 

\subsection{Example II: turning on the wormhole.}

We will now consider a transfer function with nontrivial transient effects made out of two inertial pieces connected smoothly. In particular, we assume that the transfer function is the identity in the past and a simple constant shift to a smaller $y^+$ in the future. This can be accomplished by 
\begin{equation}\label{ref:E_II}
\tau(y)=y-a\left(\frac{1+\tanh(b y)}{2}\right) \ ,
\end{equation} 
where the parameter $a$ determines the amplitude of the shift and $b$ how fast it happens around $y=0$.  It is easy to see that for negative values of $y$ we have $\tau(y)\approx y$, while for $by\gg1$ we have $\tau(y)\approx y-a$.  This means that field modes coming in from a given $y^+_{in}=y^+_0=$constant on $\mathcal{I}^-_R$ will get to  $\mathcal{I}^+_L$ along a different geodesic with $y^+_{out}=y^+_0-a$. 

\begin{figure}[h]
\centering
\includegraphics[width=0.60\textwidth]{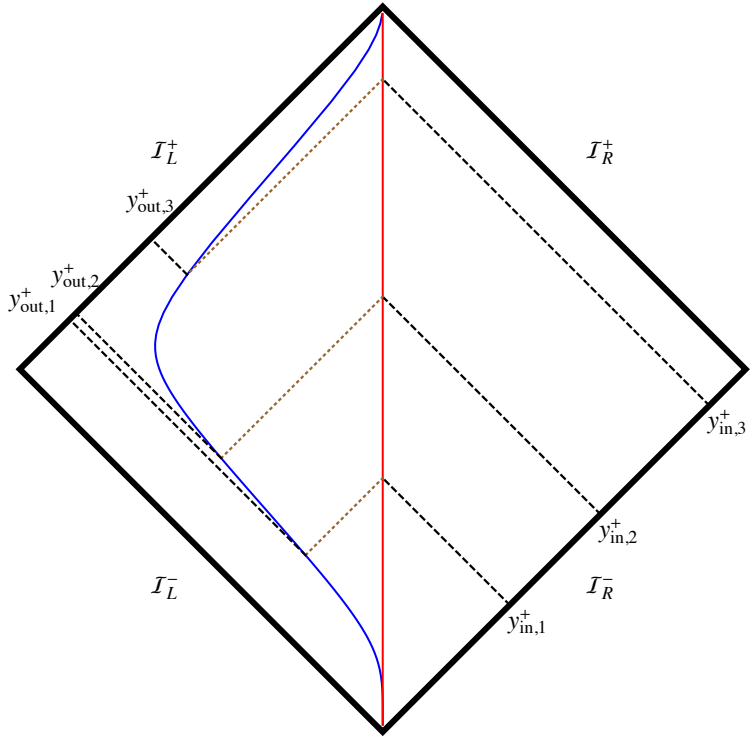}
\caption{Penrose diagram and ray tracing corresponding to the curves $\gamma_{in}(y^+)=y^+$ and $\gamma_{out}(y^+)=\tau(y^+)$, with $\tau(y^+)$ defined in Eq.(\ref{ref:E_II}). The rays are traced assuming an instantaneous transfer from $\gamma_{in}$ to $\gamma_{out}$ at a given $y^-$.  \label{fig:E_IIPD}}
\end{figure}

As we already mentioned above, the expectation value of the number operator and the energy fluxes in our 2-dimensional model are independent of the curves $\gamma_{in}$ and $\gamma_{out}$, though it is convenient to choose some specific forms for such curves in order to illustrate the process. In Fig.\ref{fig:E_IIPD} we plot $\gamma_{in}$ as the vertical line $y^+=y^-$, and $\gamma_{out}$ as the transfer function (\ref{ref:E_II}) itself, i.e., $\gamma_{out}(y^+)=\tau(y^+)$. This allows for a better visualization and interpretation of the effects of the wormhole on left-moving modes. In fact, since $\gamma_{out}$ essentially coincides with $\gamma_{in}$ for $y^+<0$, we can say that the wormhole does not operate or is under construction until it is  turned on into operation mode around $y^+\approx 0$. Around that point, the wormhole exit is smoothly separated from the entrance, which is located at $z=0$,  and taken to a new location at $z=-a$, where it is left at rest for ever after.  

The trajectory of the wormhole exit can be expressed as $z(t)$ and is determined by the relation $y^+_{out}=\tau(y^+_{in})=\tau(y^-)$. Given that $y^\pm=(t\pm z)/2$, it is easy to see that 
\begin{equation}\label{eq:zdot}
\dot{z}= \frac{\tau_{y}-1}{\tau_{y}+1} \ ,
\end{equation}
with $\tau_{y}=d\tau/dy$ and dot denoting derivative with respect to the time $t$. The fact that $|\dot{z}|$ must be smaller than unity could lead to certain constraints on the allowed transfer functions because, otherwise, the wormhole exit could travel faster than light. In the case of Eq.(\ref{ref:E_II}), we find that $\tau_{y}=1-a b/(2 \cosh^2 b y)$ requires $a b <2$ in order to guarantee $|\dot{z}|<1$ everywhere. We will explore later what happens if one saturates or tries to go beyond this bound. 

The left-moving rays in Fig.\ref{fig:E_IIPD} are drawn as follows. When a ray coming in from $\mathcal{I}^-_R$ along a geodesic $y^+$ reaches the entrance at $\gamma_{in}$, then we transfer the mode to the point $(\tau(y^+),y^+)$ on $\gamma_{out}$. Note that the image of this point of $\gamma_{out}$ on $\mathcal{I}^+_R$ coincides with the $y^-=y^+$ of the incoming mode. 

\begin{figure}[h]
\centering
\includegraphics[width=0.80\textwidth]{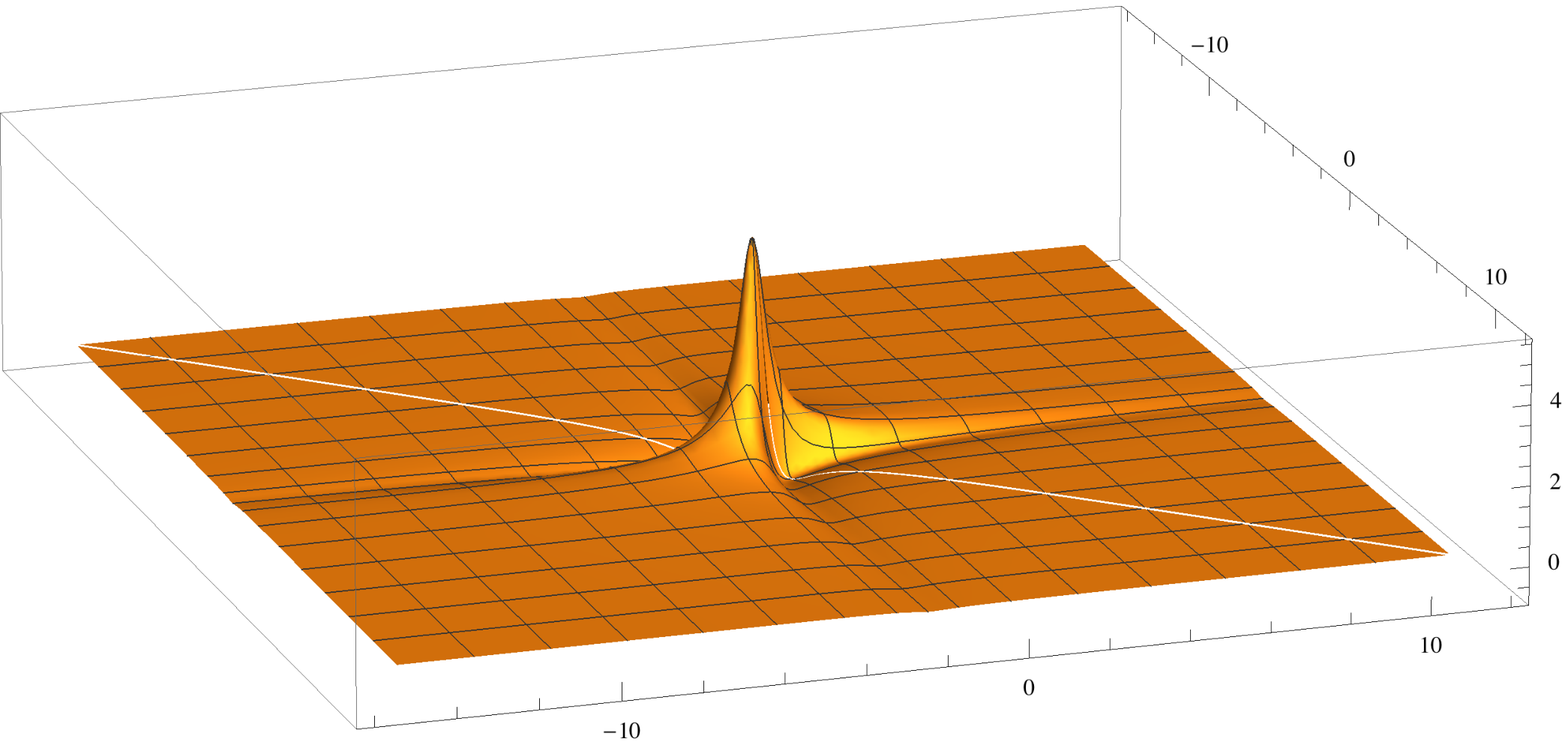}
\caption{Two-point correlation function corresponding to the transfer function of Eq.(\ref{ref:E_II}). The blank line represents the coincident points limit $y^+_1\to y^+_2$ and defines the value of the stress-energy tensor associated to the left-moving modes, $T_{++}$. The vertical axis has been magnified by a factor $10^2$ for a better visualization of the structures. \label{fig:E_IIa}}
\end{figure}

\begin{figure}[h]
\centering
\includegraphics[width=0.60\textwidth]{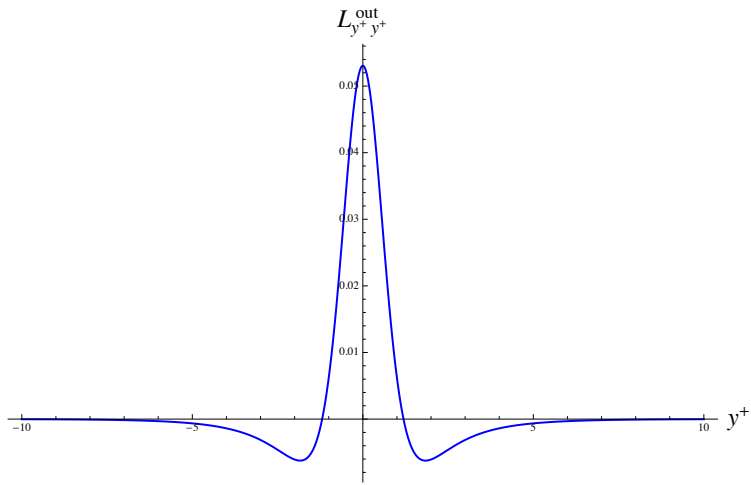}
\caption{Component $T^{out}_{++}$ of the stress-energy tensor associated to the transfer function of Eq.(\ref{ref:E_II}). In this plot we have taken $a=4$ and $b=1/3$, yielding a total energy $E\approx 0.041$. \label{fig:E_IIb}}
\end{figure}

In  Fig.(\ref{fig:E_IIa}) we represent the two-point correlator $_{X}\langle 0|:\partial_{y_1}\phi  \partial_{y_2}\phi:|0\rangle_{X}$ associated to the transfer function (\ref{ref:E_II}) evaluated in its parametric form (\ref{eq:bbxWH3}). The expectation value of the number of particles and its associated energy flux are directly related to this correlator as shown in Sec.\ref{sec:II}. The main feature of this quantity is that it is only significantly nonvanishing in the regions around $y^+\approx 0$, as shown in Fig.(\ref{fig:E_IIa}). 
If one considers wave packets instead of plane waves, it is clear that only those modes localized around $y^+=0$ will provide a nonzero contribution to the expectation value (\ref{eq:bbxWH}). The form of the stress-energy tensor appears in Fig.(\ref{fig:E_IIb}) and confirms that an energy flux appears around that same region. The total amount of radiated energy is given by the expression
\begin{equation}
E_{Tot}=\int_{-\infty}^{+\infty} T^{out}_{++}(y)dy= \frac{(3-a b (2-a b)) \sqrt{a b (2-a b)}+6 (1-a b) \tan ^{-1}\left(\frac{\sqrt{a b}}{\sqrt{2-a b}}\right)}{24 \pi  \sqrt{\frac{a}{b}} (2-a b)^{5/2}} \ ,
\end{equation}
which is real and finite as long as the product $ab<2$. Note that this is the same condition as we find above to keep $|\dot{z}|<1$. In the limit $ab\to 2$, one verifies that $E_{Tot}$ becomes negative and diverges as $\sim -1/(2-ab)^{5/2}$. Thus, as long as the exit of the wormhole moves in space-time within the light cone, the total amount of radiated energy will remain finite. Obviously, a divergent energy flux would also imply the creation of an infinite number of particles. 

\subsection{Example III: turning the wormhole on and then off.}

Now we consider a transfer funtion such that the corresponding curve $\gamma_{out}$ initially coincides with $\gamma_{in}$ then moves away from it by a certain constant spatial separation, where it stays for a while, and finally the $\gamma_{out}$ trajectory comes back to meet again with $\gamma_{in}$. Accordingly, incoming geodesics will follow $y^+=$constant trajectories initially, then will be shifted to $y^+-a$, and finally will again follow $y^+=$constant trajectories.  The corresponding transfer function can be written as
\begin{equation}\label{ref:E_III}
\tau(y)=y-a\left(\frac{\tanh(b (y-y_{on})-\tanh(c (y-y_{off})}{2}\right)  \ ,
\end{equation} 
where $y_{on}$ and $y_{off}$ denote the instants at which the wormhole is turned on and off, respectively. If we take the value of $y_{off}$ sufficiently far away form $y_{on}$, the Penrose diagram of this case would look pretty much like that in Fig.\ref{fig:E_IIPD} because of the compactification of coordinates involved, which magnifies the region around $y^+=0$ but does not allow to see the details farther away from this point. 

\begin{figure}[h]
\centering
\includegraphics[width=0.60\textwidth]{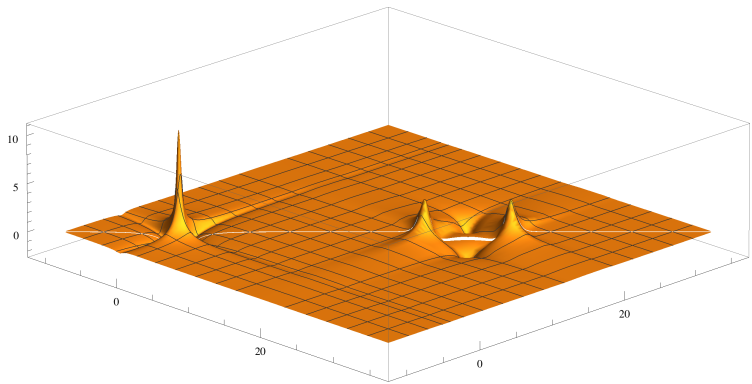}
\caption{Two-point correlation function corresponding to the transfer function of Eq.(\ref{ref:E_III}). The blank line represents the coincident points limit $y^+_1\to y^+_2$ and defines the value of the stress-energy tensor associated to the left-moving modes, $T_{++}$. The vertical axis has been magnified by a factor $2\times 10^2$ for a better visualization of the structures. \label{fig:E_IIIa}}
\end{figure}

\begin{figure}[h]
\centering
\includegraphics[width=0.60\textwidth]{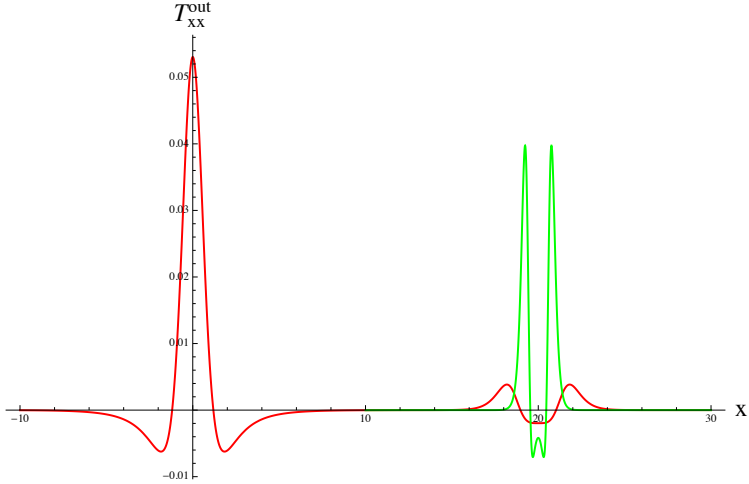}
\caption{Component $T^{out}_{++}$ of the stress-energy tensor associated to the transfer function of Eq.(\ref{ref:E_III}). In this plot we have taken $a=4$, $b=1/3$, $y_{on}=0$, and $y_{off}=20$. The red curve corresponds to $c=1$ while the green one has $c=3$. Increasing the value of $c$ leads to more pronounced peaks around $y_{off}$, though their amplitude is always bounded. \label{fig:E_IIIb}}
\end{figure}

The nonzero correlations are concentrated around those points close to the instants in which the wormhole is turned on and then off, being comparatively more intense in the initial event (see Fig. \ref{fig:E_IIIa}). The energy flux is identical to the example in Fig.(\ref{fig:E_IIb}) at the onset but has a different structure at the end, both in shape and in intensity (see Fig.(\ref{fig:E_IIIb})). Particle production will thus be sensitive to these transient periods.  The asymmetry in the on and off events can be understood by having a look at Eq.(\ref{eq:bbxWH3}) and considering the form of $\tau_X$ in each case. When the wormhole is turned into operation, we saw that $\tau_X\approx 1-ab/2\cosh^2(b(y-y_{on}))$ could become zero if $ab\ge 2$, thus causing a divergence in the two-point correlator. Such a divergence necessarily produces an infinite amount of particles and energy. 
But when the wormhole is turned off, $\tau_X\approx 1+a c/2\cosh^2(c(y-y_{off})$ cannot vanish, causing no dramatic effects. In fact, it would take $\tau_X\to \infty$ in order to get $\dot{z}=1$ but only $\tau_X\to 0$ to get $\dot{z}=-1$ (see Eq.(\ref{eq:zdot})).

\subsection{Example IV: approaching the light cone.}\label{Ex:IV}

\begin{figure}[h]
\centering
\includegraphics[width=0.60\textwidth]{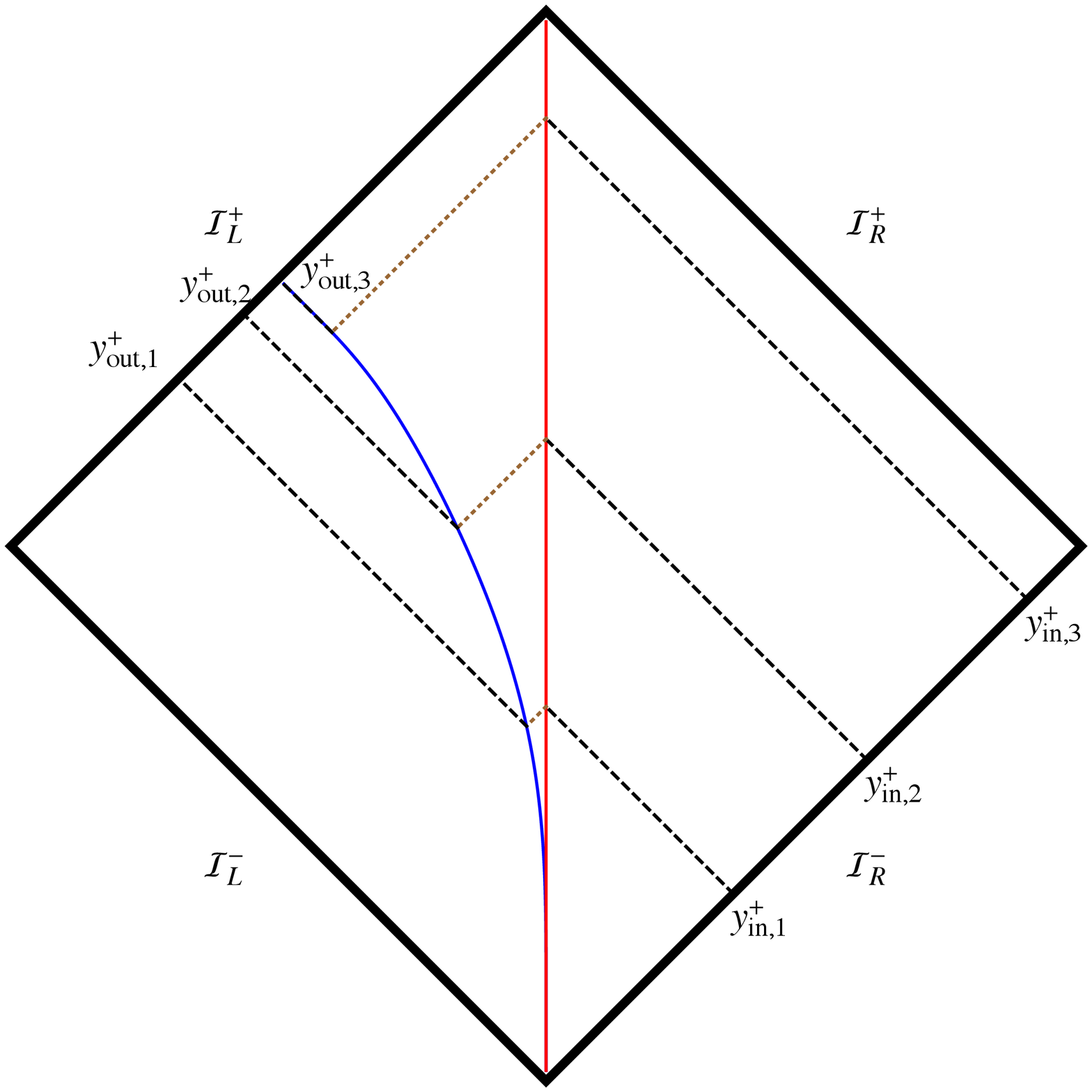}
\caption{Penrose diagram and ray tracing corresponding to the curves $\gamma_{in} \ \rightarrow \ y^+=y^-$ and $\gamma_{out} \ \rightarrow \ y^+=\tau(y^-)$, with $\tau(y)$ defined in Eq.(\ref{ref:E_IV}). The rays are traced assuming an instantaneous transfer from $\gamma_{in}$ to $\gamma_{out}$ at a given $y^-$.  \label{fig:E_IVPD}}
\end{figure}

\begin{figure}[h]
\centering
\includegraphics[width=0.60\textwidth]{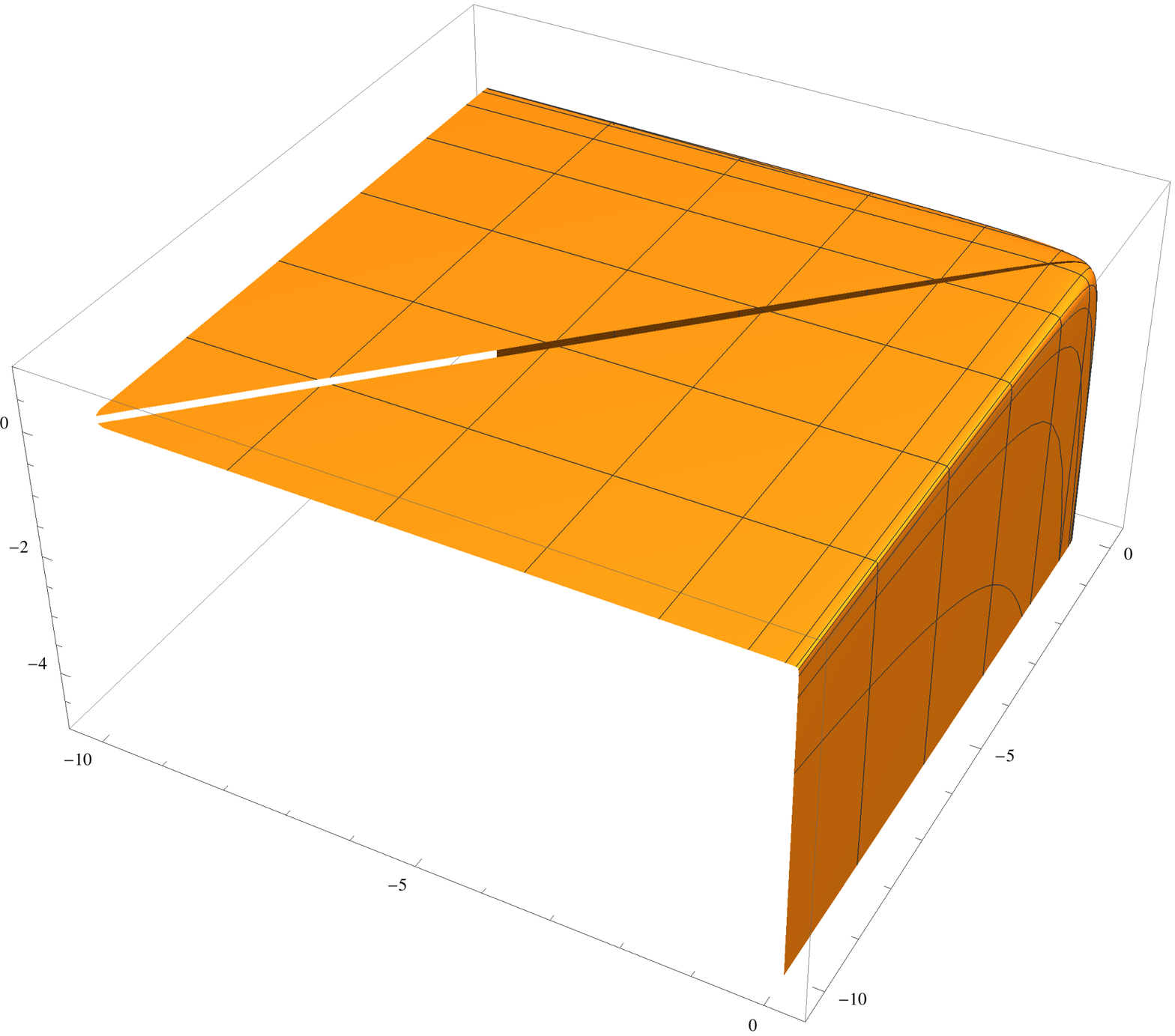}
\caption{Two-point correlation function corresponding to the transfer function of Eq.(\ref{ref:E_IV}). The blank line represents the coincident points limit $y^+_1\to y^+_2$ and defines the value of the stress-energy tensor associated to the left-moving modes, $T_{++}$. The vertical axis has been multiplied by a factor $10^{-2}$ for a better visualization of the structures. \label{fig:E_IVa}}
\end{figure}

Let us now consider that the exit of the wormhole follows an accelerated  trajectory towards the left such that it asymptotically approaches the speed of light. We take a transfer function of the form
\begin{equation}\label{ref:E_IV}
\tau(y)=\frac{y}{2}- \frac{\log[2\cosh(b y)]}{2b} , 
\end{equation}
such that for negative values of $y$ we find that $\tau_y\approx 1$ but as soon as it becomes positive, it rapidly approaches $\tau_y\approx 0$ with exponentially small corrections, which implies $\dot{z}\to -1$ for $by\gg 1$. The conformal diagram of this wormhole trajectory appears in Fig.\ref{fig:E_IVPD}. 
The two point correlation function in this case is essentially zero for $y^+_1$ and $y^+_2$ negative, and experiences an exponential drop to negative values along the boundaries $y^+_{1,2}\sim 0$, as shown in Fig. \ref{fig:E_IVa}. This fact manifests itself in the form of the stress energy tensor, which becomes $T^{out}_{++}=-\frac{b^2 e^{2 b y} \left(e^{2 b y}+2\right)}{12 \pi }$ and grows exponentially fast in the positive axis. This puts forward that this type of configuration requires an infinite amount of energy in order to be implemented (divergent negative energy flux coming from the field). An unbounded production of particles is thus expected once the exit of the wormhole is set into motion. 

\subsection{Example V: from asymptotically null to inertial trajectory.}\label{Ex:V}

\begin{figure}[h]
\centering
\includegraphics[width=0.60\textwidth]{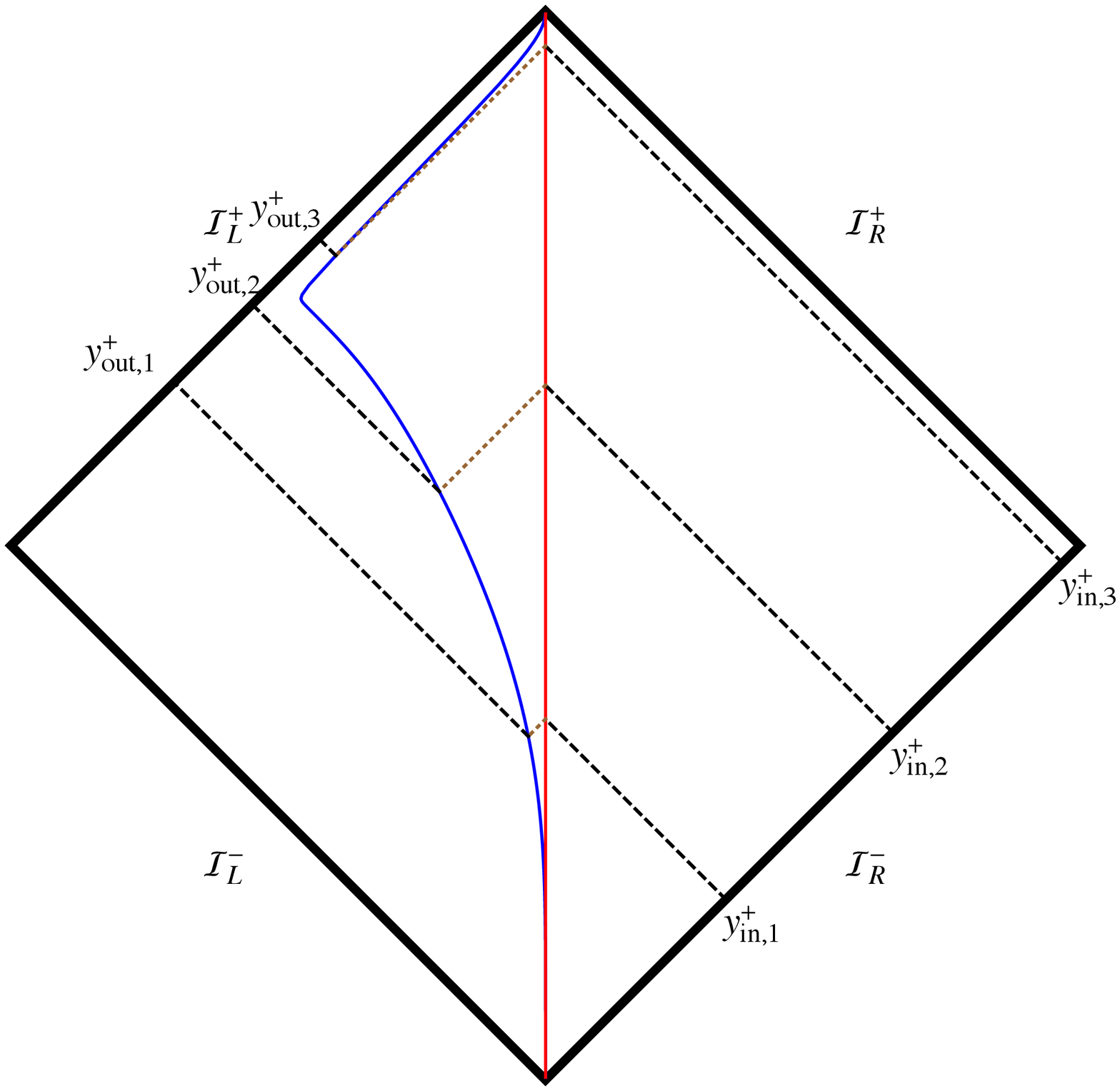}
\caption{Penrose diagram and ray tracing corresponding to the curves $\gamma_{in} \ \rightarrow \ y^+=y^-$ and $\gamma_{out} \ \rightarrow \ y^+=\tau(y^-)$, with $\tau(y)$ defined in Eq.(\ref{eq:E_V}). The rays are traced assuming an instantaneous transfer from $\gamma_{in}$ to $\gamma_{out}$ at a given $y^-$.  \label{fig:E_VPD}}
\end{figure}

\begin{figure}[h]
\centering
\includegraphics[width=0.60\textwidth]{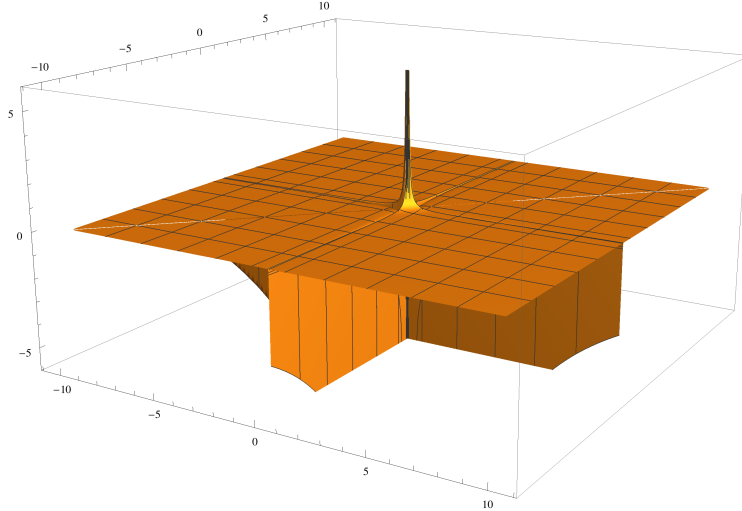}
\caption{Two-point correlation function corresponding to the transfer function of Eq.(\ref{ref:E_IV}). The blank line represents the coincident points limit $y^+_1\to y^+_2$ and defines the value of the stress-energy tensor associated to the left-moving modes, $T_{++}$. \label{fig:E_Va}}
\end{figure}

\begin{figure}[h]
\centering
\includegraphics[width=0.60\textwidth]{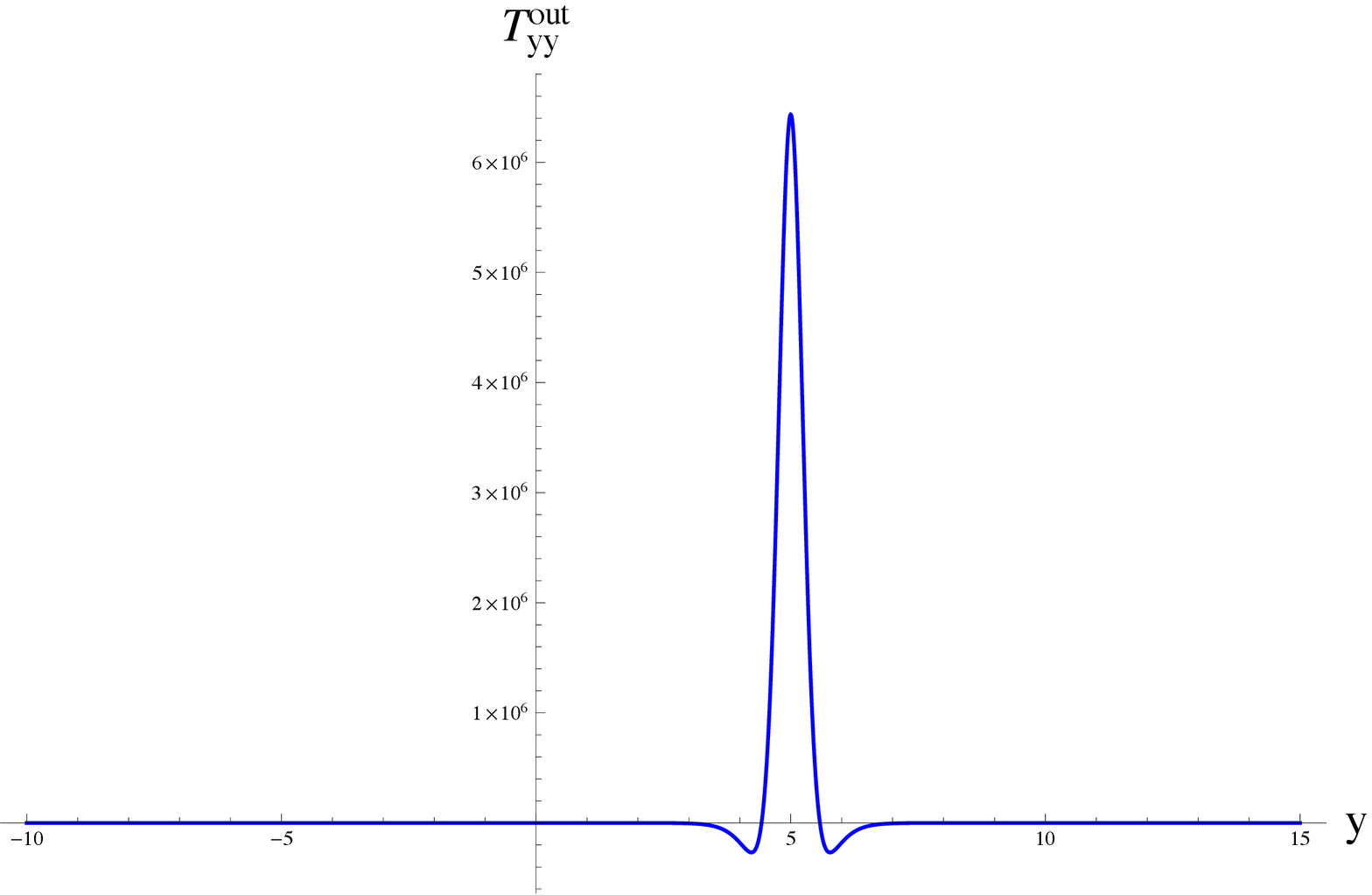}
\caption{Component $T^{out}_{++}$ of the stress-energy tensor associated to the transfer function of Eq.(\ref{eq:E_V}). In this plot we have taken $a=b=c=1$, $y_{on}=0$, and $y_{off}=10$.  \label{fig:E_Vb}}
\end{figure}

The form of the stress energy tensor of the previous example shows that the energy flux is only strictly divergent if the wormhole exit is kept accelerating forever. If the motion of the exit changes, then the exponential growth $T^{out}_{++}=-\frac{b^2 e^{2 b y} \left(e^{2 b y}+2\right)}{12 \pi }$ should be modified. This is the case we consider now. The transfer function is thus given by a piece that goes from inertial to nearly null plus another that turns the asymptotically null trajectory into an inertial one:
\begin{equation}\label{eq:E_V}
\tau(y)=\frac{y}{2}- \frac{\log[2\cosh[b (y-y_{on})]]}{2b} +\frac{y}{2}+ \frac{\log\left[\frac{\cosh[b(y-y_{off})]}{\cosh[b y_{off}]}\right]}{2b}  \ .
\end{equation}
The corresponding conformal diagram appears in Fig.\ref{fig:E_VPD}. The effect of the final inertial branch on the two-point correlator and the energy flux is remarkable, as shown in Fig.\ref{fig:E_Vb}. The two-point function in Fig. \ref{fig:E_Va} shows deep walls on the negative vertical axis around the $y_{1,2}\approx 0$ lines followed by a large positive peak. This peak provides the dominant contribution to the stress-energy tensor, and is considerably larger than the initial exponentially growing negative trend that characterizes the beginning of the asymptotically null trajectory. Interestingly, the positive peak is centered between $y_{off}$ and $y_{on}$, with symmetric negative tails on both sides. The effect of returning to an inertial trajectory is the emission of a considerable amount of particles and energy. The deep walls that appear around $y_{1,2}\approx 0$ imply the existence of long range correlations between particles generated at $y_{1,2}\approx 0$ and others generated at much later times. An indepth analysis of such correlations and their implications will be carried out elsewhere. 

\subsection{Example VI: crossing the light cone.}

\begin{figure}[h]
\centering
\includegraphics[width=0.60\textwidth]{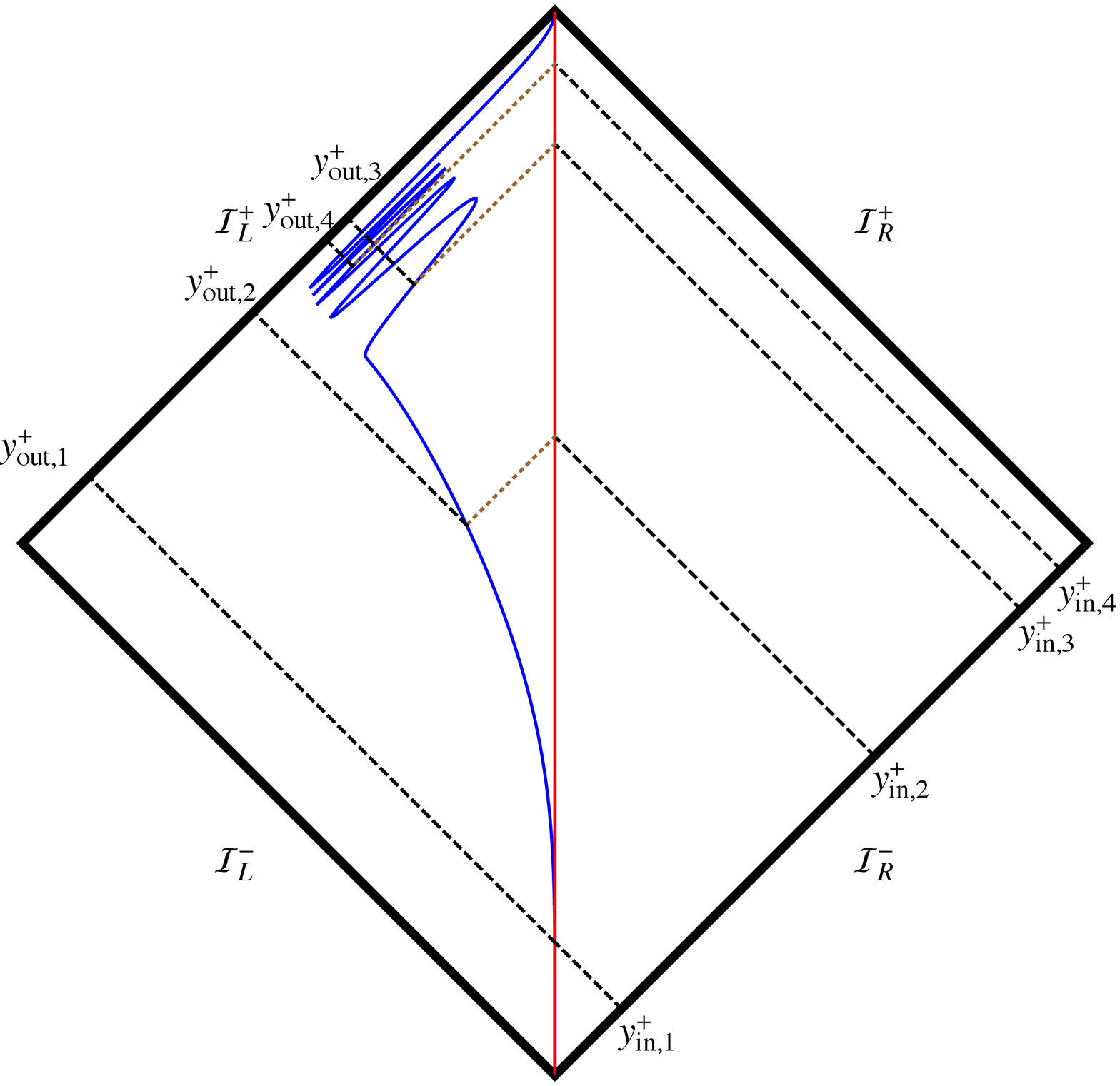}
\caption{Penrose diagram and ray tracing corresponding to the curves $\gamma_{in} \ \rightarrow \ y^+=y^-$ and $\gamma_{out} \ \rightarrow \ y^+=\tau(y^-)$, with $\tau(y)$ defined in Eq.(\ref{eq:E_V}). The rays are traced assuming an instantaneous transfer from $\gamma_{in}$ to $\gamma_{out}$ at a given $y^-$.  \label{fig:E_VIPD}}
\end{figure}

A glance at Eq.(\ref{eq:zdot}) indicates that in order to get $\dot{z}<-1$ one should have transfer functions such that  $\tau_y<0$, which necessarily requires the vanishing of $\tau_y$ at some point.  In Sec.\ref{Ex:IV} we asymptotically approached the limit $\tau_y\to 0$ but did not reach it in any finite time, and in Sec.\ref{Ex:V} we showed that if the final trajectory is relaxed into an asymptotically inertial one, then the infinite negative fluxes disappear. If one gets into the  $\tau_y<0$ region, then  divergent fluxes and particle production would be unavoidable. In Fig.\ref{fig:E_VIPD} we illustrate this effect by considering a transfer function (or wormhole exit trajectory) which oscillates around a certain interval on the $y^+_{out}$ axis as incoming modes are emitted along the $y^+_{in}$ axis. Note that the trajectory is monotonical in the $y^-$ axis. From this diagram, it is easy to see that nearby points on $\mathcal{I}^+_L$ need not necessarily be in correspondence with nearby points on  $\mathcal{I}^-_R$. As a consequence, the divergence of the two-point function $_{y}\langle 0|\partial_{y_1}\phi  \partial_{y_2}\phi|0\rangle_{y}$ as $y^+_{out,3}\to y^+_{out,4}$ may not be compensated by a similar divergence involving nearby points $y^+_{in,3}\to y^+_{in,4}$ in $_{X}\langle 0|\partial_{y_1}\phi  \partial_{y_2}\phi|0\rangle_{X}$, which causes the emergence of infinite radiation fluxes and particle production. This obstruction to crossing the lightcone indicates that using a wormhole to travel to the past is not physically allowed, which is compatible with Hawking's chronology protection conjecture. 

\section{Summary and discussion} \label{sec:I}

In this work we have presented a simplified wormhole model in $1+1$ Minkowski space-time. The model consists on a free massless scalar field that propagates subject to a boundary condition which transfers the field information from one curve $\gamma_{in}$ to another $\gamma_{out}$ via some transfer function $\tau$. The reason for this setup is that a wormhole in one spatial dimension should take particles and information from one point, the entrance, to another, the exit. This pointlike description simplifies the structural aspects of a {\it realistic} wormhole. Assuming that the entrance and the exit follow timelike trajectories, $\gamma_{in}$ and $\gamma_{out}$ respectively, besides specifying the spatial location of the entrance and the exit one must specify the instant at which the entrance is reached by the field and where, in space and time, in comes out from the exit. This is why the transfer function $\tau$ is so important. In higher spatial dimensions, one would need a similar transfer function that identifies a given section in the worldtube of the entrance with the corresponding section in the worldtube of the exit. 

The simplicity of the scalar field equation in our $1+1$-dimensional Minkowskian setup, with independent left-moving and righ-moving sectors, allows us to focus on one of the sectors only and apply an approach in terms of two-point correlation functions to explore the radiation properties of the field. We have shown that assuming total transmission of the incoming modes, the outgoing radiation is independent of the curves $\gamma_{in}$ and $\gamma_{out}$, being completely determined by the transfer function $\tau$. We have thus studied some illustrative examples to show under which circumstances particles and energy fluxes are produced. In particular, we have seen that for transfer functions of the Mobius form, $\tau(y)=\frac{a y+b}{cy+d}$ with $ad-bc\neq 0$, neither particles nor energy fluxes are produced because the vacuum state remains invariant under such (global conformal) transformations. For any other transfer function, however, particles and energy fluxes arise. 

We have illustrated this point for transformations that interpolate between different inertial branches and also in cases involving trajectories of the wormhole exit that approach the speed of light. In the former cases, the energy fluxes are finite, while in the asymptotically null cases they grow exponentially. We have also shown that if the asymptotically null transfer functions are turned into inertial ones, then the energy flux becomes bounded and positive.  As a last example, we have considered the case in which the transfer function can be seen as a trajectory involving a faster than light motion of the wormhole exit with $\dot{z}<-1$. This is equivalent to sending information to the past or to having a given event $y^+_{out}$ in correspondence with more than one event in $y^+_{in}$. This situation implies that two nearby events in $\mathcal{I}^+$ could be in correspondence with two nearby events  in $\mathcal{I}^-$ but also with two or more distant events in $\mathcal{I}^-$, thus making it impossible to cancel the divergences of the two-point functions in the coincident points limit $y^+_{out,1}\to y^+_{out,2}$ by means of similar divergences in  $\mathcal{I}^-$. Thus, unless one considers multiple $\mathcal{I}^+$ regions for a given $\mathcal{I}^-$ region such that every divergence in $y^+_{out,1}\to y^+_{out,2}$ is exactly compensated by an equivalent divergence in $y^+_{in,1}\to y^+_{in,2}$, one will be led to the generation of divergent energy fluxes. A possible way out of this problem  could be accomplished by relaxing the Hausdorff topological condition on the manifold (see chapter 19 on \cite{Visser:1995cc}). Though our analysis has been focused on a 1+1 dimensional scenario, we believe that this property of the two-point functions will be general in arbitrary dimensions due to the universal behavior of such functions at short distances for Hadamard states. Further research in this and other directions, such as the consideration of partial transmission and backscattering or the propagation of massive fields, are currently underway. \\

\section*{Acknowledgments}
G.J.O. thanks Prof. Luis Crispino and the Federal University of Par\'{a} for their kind hospitality during the elaboration of this work. G. J. O. is funded by the Ramon y Cajal contract RYC-2013-13019 (Spain).  This work is supported by the Spanish projects FIS2017-84440-C2-1-P (MINECO/FEDER, EU), SEJI/2017/042 (Generalitat Valenciana), and i-LINK1215 (CSIC).
The authors would like to acknowledge partial financial support from the European Union's Horizon 2020 research and innovation programme under the H2020-MSCA-RISE-2017 Grant No. FunFiCO-777740. Special thanks go to N. Olmo and Y. Olmo for bringing to my attention Ref. \cite{Doraemon} and for encouraging discussions on wormholes and time machines which ended up in this work.


\end{document}